\documentclass{elsart}
\usepackage{graphicx}
\usepackage{amsmath}
\usepackage{amssymb}
\usepackage{epsfig}
\usepackage{dcolumn}% Align table columns on decimal point
\usepackage{bm}% bold math
\begin{document}
\begin{frontmatter}

\title{Self-similarity degree of deformed statistical  ensembles}
\author{ A.I. Olemskoi}
\address{Institute of Applied Physics, Nat. Acad. Sci. of Ukraine \break 58, Petropavlovskaya St., 40030 Sumy,
Ukraine}
\author{ A.S. Vaylenko, I.A. Shuda}
\address{Sumy State University \break 2, Rimskii-Korsakov St., 40007 Sumy, Ukraine}

\date{}

\begin{abstract}
We consider self-similar statistical ensembles with the phase
space whose volume is invariant under the deformation that
squeezes (expands) the coordinate and expands (squeezes) the
momentum. Related probability distribution function is shown to
possess a discrete symmetry with respect to manifold action of the
Jackson derivative to be a homogeneous function with a
self-similarity degree $q$ fixed by the condition of invariance
under $(n+1)$-fold action of the dilatation operator related. In
slightly deformed phase space, we find the homogeneous function is
defined with the linear dependence at $n=0$, whereas the
self-similarity degree equals the gold mean at $n=1$, and $q\to n$
in the limit $n\to\infty$. Dilatation of the homogeneous function
is shown to decrease the self-similarity degree $q$ at $n>0$.
\end{abstract}

\end{frontmatter} \maketitle

\section{Introduction}\label{Sec.1}

Long-range interaction, long-time memory effects and evolution kinetics delayed
into power law are known to facilitate formation of multifractal phase space of
complex systems whose investigations have stipulated rise of interest to
deformed thermodynamic systems \cite{1}--\cite{4}. A cornerstone of the
deformation procedure related is formal replacement of the standard logarithm
function in the Boltzmann entropy with some deformed version. Along this line,
it is worthwhile to emphasize the Tsallis-type thermostatistics where
distribution functions are characterized with power-law tails
\cite{5}--\cite{9}, and so called basic-deformed statistical mechanics where
probability distributions have a natural cut-off in the energy spectrum
\cite{11,12}. Specific peculiarity of such type systems is
self-si\-mi\-la\-ri\-ty of related phase space whose volume is invariant under
the deformation that combines the coordinate squeezing and the momentum
expanding (or, vice versa, expanding the first axis and squeezing the second)
\cite{Vit}.

Moreover, making use of a deformation procedure in quantum mechanics allows for
to present non-trivially several physical fields from black holes to anyon
superconductivity \cite{13}. Formal basis of quantum groups related is so
called $q$-calculus \cite{Bi}--\cite{Vitiello}, originally introduced by
Heine and Jackson \cite{15,16} to study the basic hypergeometric series
\cite{gh}. From mathematical point of view, the $q$-calculus represents the
most suited formalism to investigate (multi)\-frac\-tal sets whose generation
is provided with manifold action of the dilatation operator related to the
Jackson derivative \cite{Vit}.
%A classical counterpart to the quantum group
%and $q$-deformed dynamics is based on a $q$-deformed Poisson bracket whose
%underlying algebra is invariant under action of the $q$-symplectic group
%\cite{24,25}.
It appears \cite{21a,21} for $q$-deformed bosons and fermions, a natural
generalization of the thermostatistics is based on the formalism of the
$q$-calculus, whereas stationary solution of a classical deformed Fokker-Planck
equation represents a $q$-analogue of the exponential function in the framework
of the basic hypergeometric series \cite{11,Lavagno}. Related systems exhibit a
discrete scale invariance whose description \cite{Erzan} is achieved with using
the Jackson derivative and integral, generalizing the regular derivative and
integral for discretely self-similar systems (for example, the free energy of
spin systems on a hierarchical lattice is related to a homogeneous function
being the $q$-integral \cite{Erz}).

Our consideration is devoted to determination of the exponent $q$ that plays a
central role as the non-extensivity parameter in the Tsallis thermostatistics
and as the degree of the homogeneous function in theory of self-similar
systems.\footnote{It is worthwhile to point to the notion $q$ has been
introduced in the quantum group theory \cite{QG} for the deformation parameter.
To avoid misunderstanding we will notice this parameter as $\lambda$ instead of
$q$.} Following Ref.\cite{Alex} we consider in Section \ref{Sec.2} the form of
escort probability distribution function basing on the multiplicativity
condition for this function related to a composite statistical system. In
accordance with supposition \cite{TMP}, we show the escort probability
distribution function takes the power-law form with non-integer exponent
reduced to the Tsallis non-extensivity parameter. Section \ref{Sec.3} is
devoted to short presentation of main statements of the quantum group theory
\cite{QG} to describe the symmetry of self-similar systems. Within standard
formalism of the Lee groups \cite{Jackson}, we show in Section \ref{Sec.4} that
invariance of the self-similar system under manifold action of the Jackson
derivative derives to an equation for determination of the dependence of the
self-similarity degree on the dilatation. Section \ref{Sec.5} concludes our
consideration with passage to general case of affine transformation whose
dilatation parameter changes at each step of transformation.

\section{Non-extensivity parameter}\label{Sec.2}

Formalism of non-extensive thermostatistics is known to be based on the
definition of generalized logarithm \cite{4}
\begin{equation}
\ln_q(x):=\frac{x^{1-q}-1}{1-q}\label{1}
\end{equation}
being deformed with non-extensivity parameter $q$. Making use of
this logarithm in the entropy
\begin{equation}
S:=\left<\ln_{2-q}(1/p)\right>=-\sum\limits_{i=1}^{W}\ln_{2-q}(p_i)p_i
 \label{2}
\end{equation}
accompanied with both normalization condition and definition of the internal
energy $E$
\begin{equation}
\sum\limits_{i=1}^{W}p_i=1,\qquad\sum\limits_{i=1}^{W}\varepsilon_i
\mathcal{P}\big(f(p_i)\big)=E
 \label{3}
\end{equation}
arrives at the Tsallis thermostatistics. As usually, we suppose
here to be statistical states scattered with probabilities $p_i$
and use the escort distribution
\begin{equation}
\mathcal{P}\big(f(p_i)\big):=\frac{f(p_i)}
{\sum_{i=1}^{W}f(p_i)},\qquad\sum\limits_{i=1}^{W}
\mathcal{P}\big(f(p_i)\big)=1
 \label{4}
\end{equation}
to find physical observable values type of the internal energy $E$. Within
framework of the thermostatistics version \cite{TMP}, a function $f(p_i)$ is
supposed to be of power-law:
\begin{equation}
f(p_i)\propto p_i^q.
 \label{5}
\end{equation}

It was shown recently \cite{Alex} the dependence (\ref{5}) follows from the
multiplicativity condition
\begin{equation}
f(p_{ij}^{AB})=f(p_{i}^{A})f(p_{j}^{B})
 \label{6}
\end{equation}
for composite system $AB$ consisting of non-dependent subsystems $A$ and $B$
whose probabilities are connected with the condition
$p_{ij}^{AB}=p_{i}^{A}p_{j}^{B}$. Indeed, inserting into Eq.(\ref{6}) the
series
\begin{equation}
f(x)=x^\delta\sum\limits_{n=0}^{\infty}C_n x^n
 \label{7}
\end{equation}
with unknown exponent $0<\delta<1$ and coefficients $C_n$, one derives the
equation
\begin{equation}
\sum\limits_{m=0}^{\infty}\sum\limits_{n=0}^{\infty}\left(C_m-\delta_{mn}\right)
C_n~(p_{i}^{A})^m(p_{j}^{B})^n=0.
 \label{8}
\end{equation}
As it holds for arbitrary probabilities $p_{i}^{A}$, $p_{j}^{B}$, the condition
$(C_m-\delta_{mn})C_n=0$ must be fulfilled to arrive at equalities $C_m=C_n=1$
for some fixed term $n$ of the series (\ref{7}), while $C_n=0$ otherwise.
As a result, we obtain the relation (\ref{5}) with the exponent
\begin{equation}
q=n+\delta,\quad n=0,1,2,\dots
 \label{9}
\end{equation}
which is necessary to find. Within framework of the assumption that the phase
space of non-extensive statistical system has a discrete symmetry under
deformation, we will show further the escort probability defined by Eqs.
(\ref{4}) and (\ref{5}) represents a homogeneous function with the
self-similarity degree (\ref{9}).

\section{Description of self-similarity}\label{Sec.3}

In this Section, we set forth main statements of the theory of self-similar
systems \cite{Sornette,Feder} and the quantum group theory \cite{QG} to be
needed in the following. Along this line, our consideration is based on
definition of the dilatation operator
\begin{equation}
D_x^\lambda:=\lambda^{x\partial_x},\quad\partial_x\equiv
\frac{\partial}{\partial x}.
 \label{17}
\end{equation}
With using formal expansion in the Taylor series, its action on the power-law
function gives
\begin{equation}
D_x^\lambda x^n=\lambda^{x\partial_x}x^n=\sum\limits_{m=0}^{\infty}\frac
{\left[\ln(\lambda)(x\partial_x)\right]^m}{m!}x^n=\sum\limits_{m=0}^{\infty}\frac
{\left(n\ln\lambda\right)^m}{m!}x^n=\left(\lambda x\right)^n.
 \label{power}
\end{equation}
Then, for an arbitrary analytical function $f(x)$ one obtains in a similar
manner
\begin{equation}
D_x^\lambda f(x)=\sum\limits_{m=0}^{\infty}\frac{\left[\left.\partial_x
f(x)\right|_{x=0}\right]^m}{m!}D_x^\lambda
x^m=\sum\limits_{m=0}^{\infty}\frac{\left[\left.\partial_x f(x)
\right|_{x=0}\right]^m}{m!}(\lambda x)^m=f(\lambda x).
 \label{17a}
\end{equation}
With respect to a homogeneous function defined with the condition
\begin{equation}
h(\lambda x):=\lambda^ q h(x)
 \label{hom}
\end{equation}
action of the $\lambda$-dilatation operator (\ref{17}) is described by the
eigen value equation
\begin{equation}
D_\lambda h(x)=\lambda^ q h(x).
 \label{eigen}
\end{equation}
On the other hand, the Jackson derivative
\begin{equation}
\mathcal{D}_x^\lambda:=\frac{D_x^\lambda-1}{(\lambda-1)x},
 \label{18}
\end{equation}
connected with the dilatation operator by means of the commutator
$[\mathcal{D}_x^\lambda,x]=D_x^\lambda$, is characterized by the equality
\begin{equation}
\big(x\mathcal{D}_x^\lambda\big)h(x)=[q]_\lambda h(x).
 \label{19}
\end{equation}
Here, $h(x)$ is the homogeneous function subject the self-similarity property
(\ref{hom}) and the $\lambda$-basic number
\begin{equation}
[q]_\lambda:=\frac{\lambda^q-1}{\lambda-1}
 \label{20}
\end{equation}
is introduced to be equal the exponent $q$ in the absence of dilatation
$(\lambda=1)$ and to increase as $\lambda^{q-1}$ in the limit
$\lambda\to\infty$. According to Eqs. (\ref{eigen}) and (\ref{19}) the
homogeneous function $h(x)$ is eigen function of both $\lambda$-dilatation
operator $D_\lambda$ and $\lambda$-derivative $x\mathcal{D}_x^\lambda$ whose
eigen-values are the power-law function $\lambda^q$ and $\lambda$-basic number
$[q]_\lambda$ given by Eq.(\ref{20}), respectively.

With using the $\lambda$-deformed Leibnitz rule for some functions $f(x)$ and
$g(x)$ \cite{gh}
\begin{equation} \label{18a}
\begin{split}
\mathcal{D}_x^\lambda f(x)g(x)=g(x)\mathcal{D}_x^\lambda f(x)+ f(\lambda
x)\mathcal{D}_x^\lambda g(x)\\=g(\lambda x)\mathcal{D}_x^\lambda f(x)+
f(x)\mathcal{D}_x^\lambda g(x),
\end{split}
\end{equation}
it is easy to convince the homogeneous function, being the solution of the
eigen equation (\ref{19}), has the form
\begin{equation}
h(x)=A_\lambda(x)x^q
 \label{21a}
\end{equation}
to be determined with a degree $q>0$ and an amplitude $A_\lambda(x)$ obeying
the condition
\begin{equation}
{\mathcal{D}^\lambda_x}A_\lambda(x)=0.
 \label{21b}
\end{equation}
From here, with accounting for Eqs. (\ref{18}) and (\ref{17a}) one obtains the
$\lambda$-periodicity property
\begin{equation}
A_\lambda(\lambda x)=A_\lambda(x).
 \label{21c}
\end{equation}
This property is easy to show to be satisfied with the series
\begin{equation}
A(x)=x^{-q}\sum\limits_{m=-\infty}^{\infty}p(\lambda^m x)\lambda^{-qm}
 \label{21s}
\end{equation}
where $p(x)$ is arbitrary periodic function vanishing at $x=0$ together with
its first $n+1$ derivatives where $n\equiv[q]$ is integer of $q$. According to
Eq.(\ref{21s}) the amplitude of the homogeneous function (\ref{21a}) is
periodical in $\ln x$ with period $\ln\lambda$ \cite{Erzan}. In the simplest
case $p(x)=1-\cos(x)$, one obtains the Weierstrass-Mandelbrot function
\begin{equation}
h(x)=\sum\limits_{m=-\infty}^{\infty}\frac{1-\cos(\lambda^m x)}{\lambda^{qm}}
 \label{21ss}
\end{equation}
whose graph is a fractal set with the dimension $2-q$ \cite{Feder}. Introducing
the self-similarity exponent in complex form
\begin{equation}
{\rm q}_m\equiv q+{\rm i}\frac{2\pi}{\ln\lambda}m,\qquad m=0,\pm 1,\pm 2,\dots,
 \label{21e}
\end{equation}
one represents the homogeneous function (\ref{21a}) as the Melline series
\begin{equation}
h(x)=\sum\limits_{m=-\infty}^{\infty}A_mx^{{\rm q}_m}
 \label{21f}
\end{equation}
to be the formal hallmark of the self-similarity. However, it is much more
convenient physically to use the Fourier representation \cite{Sor}
\begin{equation}
A_\lambda(x)=\sum\limits_{m=-\infty}^{\infty}A_m\exp\left({\rm
i}\frac{2\pi}{\ln\lambda}m\ln x\right)\simeq A_0+2A_1\cos\left(2\pi\frac{\ln
x}{\ln\lambda}\right)
 \label{21d}
\end{equation}
where the last estimation takes into account that the coefficients $A_m=A_{-m}$
decay very fast with the $m$ increase. Logarithmically oscillating behavior is
known \cite{Ol} to relate to increase of levels number of hierarchical tree
which describes a dilatation cascade under manifold deformation of the
self-similar system.

Thus, above formalism describes properties of statistical systems possessing
invariance under discrete deformations with a finite scale \cite{Sor}. The
phase space of such systems is known to form a fractal set whose generation is
provided with manifold action of the dilatation operator (\ref{17}) that
expands $(\lambda>1)$ or squeezes $(\lambda<1)$ the $x$ axis with the finite
scale $\lambda$. Along this line, the set $\{h(x)\}$ of homogeneous functions
(\ref{21a}) forms a vector space which provides the Fock-Bargmann
representation $\{\left|\psi\right>\}$ of the Weyl-Heisenberg algebra generated
by the set of creation and annihilation operators $\hat{a}$, $a$ accompanied
with the particle number operator $N\equiv\hat{a}a$ to be obeyed the
commutation relations \cite{QG,Perelomov}
\begin{equation}
[a,\hat{a}]=1,\quad[N,\hat{a}]=\hat{a},\quad[a,N]=a.
 \label{21ga}
\end{equation}
The passage from the Hilbert space $\{\left|\psi\right>\}$ to the functional
one $\{h(x)\}$ is provided with the following identification:
\begin{equation}
\hat{a}\to x,\quad a\to\partial_x,\quad N\to x\partial_x.
 \label{21gb}
\end{equation}
Remarkably, the homogeneous functions are marked out among the whole set of
functions in a similar manner as coherent states make the same among all
possible quantum states \cite{Vitiello,Vit}. In fact, the wave function related
to the coherent states
\begin{eqnarray} \label{21gc}
\left|x\right>&:=\exp\left(x\hat{a}-\bar{x}a\right)\left|0\right>
=\exp\left(-\frac{|x|^2}{2}\right)\sum\limits_{n=0}^{\infty}
\frac{\left(x\hat{a}\right)^n}{n!}\left|0\right>\nonumber\\
&=\exp\left(-\frac{|x|^2}{2}\right)\sum\limits_{n=0}^{\infty}
h_n(x)\left|n\right>\\ \nonumber
\end{eqnarray}
is reduced to expansion over eigenkets
$\left|n\right>=\frac{1}{\sqrt{n!}}\hat{a}^n\left|0\right>$ with coefficients
$h_n(x)\equiv\left.h(x)\right|_{q=n}$ being the homogeneous function
(\ref{21a}) with the integer self-similarity degree $q=n$. In accordance with
Eq.(\ref{17a}), one obtains the deformed coherent state
\begin{equation}
D_x^\lambda\left|x\right>=\left|\lambda x\right>=\exp\left(-\frac{|\lambda
x|^2}{2}\right)\sum\limits_{n=0}^{\infty}\frac{(\lambda
x)^n}{\sqrt{n!}}\left|n\right>
 \label{21gac}
\end{equation}
whose comparison with the last of expressions (\ref{21gc}) shows that, with
accuracy to non-essential multiplier $1/\sqrt{n!}$, the deformed homogeneous
function
\begin{equation}
\tilde{h}_n(x):=D_x^\lambda h_n(x)=\frac{(\lambda x)^n}{\sqrt{n!}}
 \label{21gbc}
\end{equation}
related to fractal set differs from the non-deformed one by the factor
$\lambda^n$ being inherent in the squeezing transformation in the course of the
fractal generation \cite{Feder}. Thus, we can conclude that manifold dilatation
of the coherent state (\ref{21gc}) relates to the fractal iteration process.

With deformation of the Weyl-Heisenberg algebra (\ref{21ga}), the operator $a$
in relations (\ref{21gb}) must be replaced with the Jackson derivative
(\ref{18}). Moreover, similar to the quantum optics, it is convenient to
introduce the squeezing operator \cite{Vitiello,Vit}
\begin{equation}
S:=\sqrt{\lambda}~D_x^\lambda=\sqrt{\lambda}~{\rm e }^{(\ln\lambda)N},\quad
N\equiv x\partial_x
 \label{a}
\end{equation}
that picks out the factor $\sqrt{\lambda}$ in front of the dilatation operator
(\ref{17}). At real $x$ values, the coordinate $x$ and the momentum $p=-{\rm
i}\partial_x$ are associated to the annihilation and creation operators
\begin{equation}
c:=\frac{1}{\sqrt{2}}\left(\hat{a}+a\right),\quad
c^\dag:=\frac{1}{\sqrt{2}}\left(\hat{a}-a\right)
 \label{b}
\end{equation}
being hermitian conjugated. At the squeezing, these operators transform as
follows \cite{Vit}
\begin{equation} \label{c}
\begin{split}
\tilde{c}:=S^{-1}cS= \frac{1}{\sqrt{2}}\left(\lambda^{-1}\hat{a}+\lambda
a\right)=\cosh(\ln\lambda)c-\sinh(\ln\lambda)c^\dag,\\ \tilde{c}^\dag
:=S^{-1}c^\dag S= \frac{1}{\sqrt{2}}\left(\lambda^{-1}\hat{a}-\lambda
a\right)=\cosh(\ln\lambda)c^\dag -\sinh(\ln\lambda)c.
\end{split}
\end{equation}
Above relations show the squeezing transformation is reduced to the Bogoliubov
rotation with the angle $\ln\lambda$ in a plane spanned by the operators $c$
and $c^\dag $. This means the related transformations of the pair of conjugated
coordinate and momentum
\begin{equation}
\tilde{x}:=S^{-1}xS=\lambda^{-1}x,\quad \tilde{p}:=S^{-1}pS=\lambda p
 \label{d}
\end{equation}
keep the phase space volume to be invariant under the squeezing.

The measure of a fractal set generated by the dilatation transformation
\cite{Erzan}
\begin{equation}
M(\ell_m,R)=A_\lambda(\ell_m,R)\ell_m^d(\ell_m/R)^{-d_f}
 \label{21g}
\end{equation}
depends in the power form of the linear size $\ell_m=\lambda^m$ of covering
balls related to $m$-th step of the fractal generation. (Here,
$A_\lambda(\ell_m,R)$ is an oscillation amplitude type of $A_\lambda(x)$ in the
homogeneous function (\ref{21a}), $R$ is characteristic size of a fractal
region of dimension $d_f$ embedded in a space with topological dimension $d\geq
d_f$.) According to Eq.(\ref{17a}) the dilatation (\ref{17}) transforms a box
of size $\ell_m$ into the same of the size $\ell_{m+1}=\lambda\ell_m$, so that
one has
\begin{equation}
D_\ell^\lambda M(\ell_m,R)=M(\ell_{m+1},R)=\lambda^{d-d_f}M(\ell_m,R).
 \label{21h}
\end{equation}
In the limit $m\to\infty$, the measure $M(R):=\lim_{m\to\infty}M(\ell_m,R)$
depends only of the characteristic fractal size $R$, for which the operator
$D_R^\lambda$ gives the eigen value equation
\begin{equation}
D_R^\lambda M(R)=R^{d_f}M(R).
 \label{21i}
\end{equation}
Thus, the generator (\ref{17}) of the dilatation group allows for to find the
difference $d-d_f$ between topological and fractal dimensions -- if the scale
$\ell$ is dilated, and fractal dimension $d_f$ itself -- at dilating the
fractal size $R$. On the other hand, making use of combinations
$\ell\mathcal{D}_\ell^\lambda$ and $R\mathcal{D}_R^\lambda$ with the Jackson
derivative (\ref{18}) arrives at the $\lambda$-basic numbers $[d-d_f]_\lambda$
and $[d_f]_\lambda$, respectively.

In more complicated case, when complex system has a multifractal space
\cite{Sor,OlKhB}, its measure is characterized by the partition function
\begin{equation}
Z_\lambda(\ell,R;q):=\sum\limits_{i}^{N(\ell,R)}\ell^{\alpha q}
 \label{21j}
\end{equation}
where summation is carried out over all balls $i$ whose number
$N:=\ell^{-f(\alpha)}$ is determined by a spectral function $f(\alpha)$ with
argument being a singularity strength $\alpha$ \cite{multi}. Similarly to
Eq.(\ref{21g}), one obtains the homogeneous function
\begin{equation}
Z_\lambda(\ell,R;q)=A_\lambda(\ell,R;q)(\ell/R)^{\tau(q)}
 \label{21k}
\end{equation}
where $A_\lambda(\ell,R;q)$ is an oscillating amplitude and $\tau(q)$ is the
mass exponent defined by the equation
\begin{equation}
\tau(q)=\alpha_q q-f(\alpha_q)
 \label{21l}
\end{equation}
with the specific singularity strength $\alpha_q$ being fixed by the conditions
of the steepest-descent method:
\begin{equation}
\left.\frac{{\rm d}f}{{\rm d}\alpha}\right|_{\alpha=\alpha_q}=q,\quad
\left.\frac{{\rm d}^2 f}{{\rm d}\alpha^2}\right|_{\alpha=\alpha_q}<0.
 \label{21m}
\end{equation}
As a result, the equality type of (\ref{21i})
\begin{equation}
D_\ell^\lambda Z_\lambda(\ell,R;q)=\lambda^{\tau(q)}Z_\lambda(\ell,R;q)
 \label{21n}
\end{equation}
determines the mass exponent $\tau(q)$. Analogously, making use of the
differentiation operator $\ell\mathcal{D}_\ell^\lambda$ gives the
$\lambda$-basic number $[\tau(q)]_\lambda$.

\section{Determining the self-similarity degree}\label{Sec.4}

Within the Lee group theory \cite{Jackson}, the operator
$x\mathcal{D}_x^\lambda$ appears as a generator of the transformation
\begin{equation}
T_x^\lambda(t):=\exp[t(x\mathcal{D}_x^\lambda)]
 \label{17aff}
\end{equation}
related to a continuous parameter $t$. With accounting for Eq.(\ref{19}), the
formal expansion of the exponential operator (\ref{17aff}) in the Taylor series
gives
\begin{equation} \label{17af}
T_x^\lambda(t)h(x)=\sum_{m=0}^{\infty}\frac{t^m(x\mathcal{D}_x^\lambda)^m}{m!}h(x)
=\sum_{m=0}^{\infty}\frac{([q]_\lambda t)^m}{m!}h(x)={\rm e}^{[q]_\lambda
t}h(x).
\end{equation}
Thus, the group operator (\ref{17aff}) has the eigen value ${\rm
e}^{[q]_\lambda t}$ determined by the $\lambda$-basic number (\ref{20}) and the
eigen function being the homogeneous function (\ref{21a}).

Let us introduce now the definition of the Lee group transformation
\begin{equation}
\mathcal{T}_x^\lambda(t):=\exp(t\mathcal{D}_x^\lambda)
 \label{17afg}
\end{equation}
being slightly different from Eq.(\ref{17aff}) because the Jackson derivative
itself $\mathcal{D}_x^\lambda$ is taken as the generator instead of the
combination $x\mathcal{D}_x^\lambda$. Our main approach is based on the
proposition the self-similar system is invariant under the $(n+1)$-fold action
of the $\lambda$-differentiation operator\footnote{It is interesting to point
out a periodic function $p(x)$ in the series (\ref{21s}) has as well $n+1$
first derivatives vanishing in the point $x=0$.}
\begin{equation}
\left({\mathcal{D}^\lambda_x}\right)^{n+1}h(x):=x^{-(n+1)}h(x).
 \label{21c}
\end{equation}
Inserting here related homogeneous function (\ref{21a}), with accounting for
Eq.(\ref{21b}) one obtains
\begin{equation}
\left({\mathcal{D}^\lambda_x}\right)^{n+1}x^{q}=\left(\prod_{m=0}^{n}[q-m]_\lambda\right)
x^{q-(n+1)}.
 \label{21}
\end{equation}
Then, the self-similarity condition (\ref{21c}) gives the transcendental
equation
\begin{equation}
\prod_{m=0}^{n}[q-m]_\lambda=1.
 \label{22}
\end{equation}
As a result, similarly to Eqs.(\ref{17af}), the action of the dilatation group
operator (\ref{17afg}) is defined with chain of the following equalities:
\begin{equation} \label{17ag}
\begin{split}
\mathcal{T}_x^\lambda(t)h(x)=\left[1+\sum_{m=0}^{\infty}
\frac{t^{m+1}(\mathcal{D}_x^\lambda)^{m+1}}{(m+1)!}\right]h(x)
\\=\left[1+\sum_{m=0}^{\infty}\frac{\prod_{l=0}^{m}[q-l]_\lambda
~t^{m+1}x^{-(m+1)}}{(m+1)!}\right]h(x)
=\sum_{m=0}^{\infty}\frac{(t/x)^m}{m!}h(x)={\rm e}^{t/x}h(x).
\end{split}
\end{equation}
Comparison of Eqs. (\ref{17af}) and (\ref{17ag}) at $t=1$ shows the generator
$x\mathcal{D}_x^\lambda$ creates the exponential eigen value with the basic
number $[q]_\lambda$, whereas making use of the Jackson derivative
$\mathcal{D}_x^\lambda$, accompanied with the self-similarity condition
(\ref{22}), makes it with the exponent $1/x$ being the inverse argument of the
homogeneous function. Physically, result (\ref{17ag}) means the dilatation
strengthens exponentially the power-type function (\ref{21a}) within the domain
of small values $x$. Principally important is that this property is provided
with the self-similarity condition (\ref{22}) only.

\begin{figure}
\centering
 {\includegraphics[width=135mm]{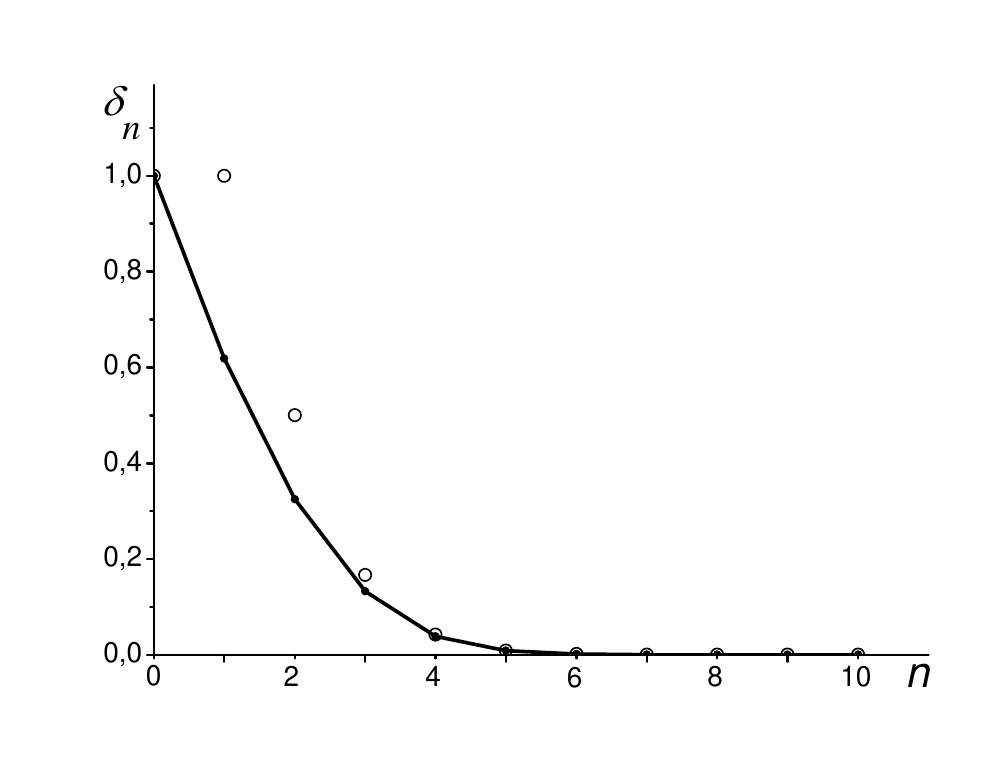}}
 {\includegraphics[width=135mm]{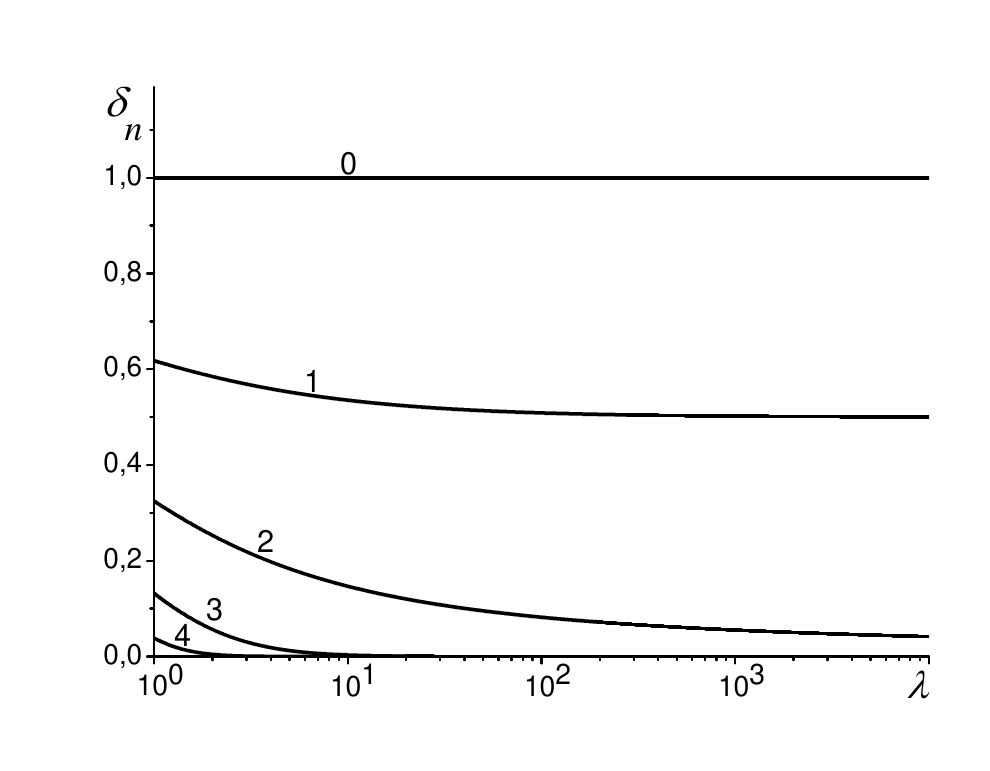}}
 \caption{Top: Dependence of the relative self-similarity degree $\delta_n=q_n-n$
on the number $n+1$ of the dilatation steps in slightly dilated system,
$\lambda\to1$ [points connected with solid line show numerical solutions of
Eq.(\ref{22a}), open cycles relate to analytical solution (\ref{27a})]; bottom:
the same in dependence of the dilatation parameter $\lambda$ at different
integers $n$ being pointed out near related curves.}\label{fig.1}
\end{figure}

As a result, one needs to study when this condition is fulfilled. In slightly
dilated system $(\lambda\to 1)$ the condition (\ref{22}) takes the simple form
\begin{equation}
\prod_{m=0}^{n}(q-m)=1.
 \label{22a}
\end{equation}
In the case of one-fold dilatation $(n=0)$, the self-similarity degree $q_0=1$
relates to the linear function (\ref{21a}). Much more interesting situation is
realized at two-fold deformation $(n=1)$, when the self-similarity degree is
reduced to {\it the gold mean}
\begin{equation}
 q_1=\frac{1+\sqrt{5}}{2}\simeq1.618.
 \label{23a}
\end{equation}
For dilatation orders $n>1$, one obtains more complicated expressions, which is
convenient to present in graphical form (see Fig.\ref{fig.1}a).
This figure shows the self-similarity degree can be expressed as the binomial
(\ref{9}) that defines the non-extensivity parameter with fractional addition
$0<\delta<1$. Within the first order of accuracy over $\delta$, Eq.(\ref{22a})
takes the simple form
\begin{equation}
n!\delta_n\approx 1.
 \label{24a}
\end{equation}
As a result, the self-similarity degree of the slightly dilated system reads
\begin{equation}
 q_n\approx n+\frac{1}{n!}.
 \label{27a}
\end{equation}
Fig.\ref{fig.1}a shows this expression at $n=1$ gives maximum error 47\% which
falls down very quickly achieving values less than 1\% at $n>3$.

In general case of the $\lambda$-dilated systems, combination of Eqs.
(\ref{22}), (\ref{20}) and (\ref{9}) arrives at the equality
\begin{equation}
\delta_n\approx\frac{1}{\ln
\lambda}\ln\left(1+\frac{\lambda-1}{[n]_\lambda!}\right)
 \label{26}
\end{equation}
and the self-similarity degree takes the estimation
\begin{equation}
q_n\approx n+\frac{1}{\ln
\lambda}\ln\left(1+\frac{\lambda-1}{[n]_\lambda!}\right).
 \label{27}
\end{equation}
Corresponding values $\delta_n=q_n-n$ at different positive integers $n$ are
shown in Fig.\ref{fig.1}b. At $n=0$, the self-similarity degree $q_0=1$ does
not depend on the dilatation $\lambda$, whereas with the $n>0$ increase the
$q_n$ value monotonically decays the faster the more the dilatation order $n$.
At that, the expression (\ref{27}) gives error not more than 1\% at $n>3$.

\section{Discussion}\label{Sec.5}

Above consideration concerns the properties of the self-similar statistical
ensembles whose phase space has invariant volume since related deformations
(\ref{d}) combine coordinate squeezing with equal momentum expanding (and vice
ver\-sa). According to Eq.(\ref{17a}), the mathematical tool to describe the
deformation is based on the definition (\ref{17}) of the dilatation operator
accompanied with the Jackson derivative (\ref{18}). In addition, the
probability distribution function is reduced to the homogeneous function
(\ref{21a}) being the eigen function of both dilatation operator and Jackson
derivative whose eigen-values $\lambda^q$ and $[q]_\lambda$ are determined by
the self-similarity degree $q$. On the other hand, statistical states of
self-similar ensemble form a (multi)fractal set whose generation relates to
manifold dilatations (\ref{21gbc}) of the homogeneous function with the integer
self-similarity degree. As show Eqs. (\ref{21i}) and (\ref{21n}), making use of
the Jackson derivative allows for to find the fractal dimension $d_f$ as well
as the mass exponent $\tau(q)$ of corresponding fractal and multifractal sets.

As shows the comparison of Sections \ref{Sec.2} and \ref{Sec.3}, the
non-extensivity parameter (\ref{9}) of the Tsallis thermostatistics coincides
with the self-similarity degree $q$ of the homogeneous function (\ref{21a}).
Being the probability distribution, this function is supposed under the
manifold action (\ref{21c}) of the Jackson derivative to be multiplied by the
trivial power-law factor without the product (\ref{22}) being put to equal one.
Then, the action (\ref{17ag}) of the Lee group operator (\ref{17afg}) picks out
the exponential factor which sharpens the power-law form of the homogeneous
function within the domain of small argument values. As a result, we derive the
equation (\ref{22}) which defines the dependence of the self-similarity degree
$q$ on the dilatation $\lambda$ at different numbers $n+1$ of deformation
iterations.

In the limit $\lambda\to1$ related to the slightly deformed systems, we find
the homogeneous function (\ref{21a}) is defined with the linear dependence
$(q_0=1)$ at the one-fold deformation $(n=0)$. Principle different case takes
place at two-fold deformation $(n=1)$ when the self-similarity degree is
reduced to the gold mean (\ref{23a}) being a hallmark of self-similarity
\cite{Ol}. According to the estimation (\ref{27a}), at $n>3$, the degree $q_n$
is slightly different of its integer $[q]$ being equal the number $n$.

As shows Fig.\ref{fig.1}b, a finite dilatation $\lambda\ne1$ of a self-similar
statistical ensemble keeps invariant the linear form of the homogeneous
function at the one-fold deformation $(n=0)$. Increase of the number $n>0$
arrives at dependencies $q_n(\lambda)$ that decay logarithmically slowly with
the dilatation growth. In the limit $n\to\infty$, this decaying is estimated
with the relation (\ref{27}).

Touching on the physical meaning of the results obtained, we emphasize that
above used self-similarity transformations are characterized with unique
dilatation parameter $\lambda$ being constant at all steps of transformations.
In general case of {\it affine} transformations, one needs to consider a set of
dilatation parameters $\{\lambda_m\}$ which can change at each step $m$ of
transformations. Moreover, in statistical systems, each of these
transformations gets a specific weight $\alpha_m\geq 0$ normalized with the
condition $\sum_m\alpha_m=1$. As a result, the dilatation operator (\ref{17})
takes the form
\begin{equation}
D_x:=\sum\limits_{m=-\infty}^{\infty}\alpha_mD_x^{\lambda_m}=
\sum\limits_{m=-\infty}^{\infty}\alpha_m\lambda_m^{x\partial_x}
 \label{17d}
\end{equation}
to be related to the affine transformation
\begin{equation}
D_xf(x)=\sum\limits_{m=-\infty}^{\infty}\alpha_m f(\lambda_m x)
 \label{17ad}
\end{equation}
instead of the homogeneous dilatation (\ref{17a}).

Consideration of self-affine systems whose observables are invariant under
transformations (\ref{17ad}) shows the inverse Melline transformation of
coordinate dependence of such an observable has poles on the complex plane $z$
at condition (see \cite{Makarov}, Appendix in the last of Refs.\cite{Sor} or
Section 3.2 in Ref.\cite{Ol})
\begin{equation}
\sum\limits_{m=-\infty}^{\infty}\alpha_m^2\lambda_m^{d_f+{\rm i}z}=1.
 \label{ed}
\end{equation}
These poles are ranged in rows parallel to real axis to be located within the
band $1-d_f\leq\Im{z}\leq0$ whose width is fixed by the fractal dimension $d_f$
of self-affine set obtained. The solutions of the equation (\ref{ed})
\begin{equation}
z_n^m=-{\rm i}\zeta_n+\frac{2\pi}{\ln\lambda_n}m,\quad m=0,\pm 1,\pm 2,\dots
 \label{21ed}
\end{equation}
are determined with set of parameters $\lambda_n\geq1$ and $0\leq\zeta_n\leq
d_f$ related to the pole rows $n=0,1,\dots,\nu$ (integer $m$ defines the number
of pole in given row). Respectively, the amplitude (\ref{21s}) of the
homogeneous function (\ref{21a}) takes the form
\begin{equation}
A(x)=\sum\limits_{n=0}^\nu x^{-\zeta_n}p_n\left(\frac{\ln
x}{\ln\lambda_n}\right).
 \label{aada}
\end{equation}
In difference of self-similar systems, the amplitude (\ref{aada}) of
self-affine homogeneous function does not only oscillate logarithmically, but
decays monotonically with argument increase if the number of pole rows is
$\nu+1>1$.

In trivial case of simple systems, the phase space is smooth so that the
fractal dimension $d_f=1$ and all poles (\ref{21ed}) are pure real. In this
case, the values $\alpha_m=\lambda_m^{-1}=N^{-1}$ are determined by total
number $N\to\infty$ of steps in the fractal generating, and above poles read
\begin{equation}
z_0^m=\frac{2\pi}{\ln N}m,\quad m=0,\pm 1,\pm 2,\dots
 \label{21eda}
\end{equation}
Then, the homogeneous function (\ref{21a}) is reduced to the linear power law
$(q=1)$ with logarithmically periodical amplitude
\begin{equation}
p_0(x)=\sum\limits_{m=-\infty}^{\infty}\alpha_mp_0\left(m~\frac{\ln x}{\ln N
}\right).
 \label{abda}
\end{equation}

As pointed out in Introduction, long-range interaction and long-time memory
effects inherent in complex systems facilitate formation of fractal phase space
with dimension $d_f<1$. Strengthening these effects decreases the fractal
dimension to arrive at creation and consequent increase of imaginary pole rows
in the points (\ref{21ed}) with $n>0$. One can suppose these poles relate to
different numbers $n$ of manifold deformation (\ref{21c}) that determines the
self-similarity degree $q_n$ in dependence of the dilatation. In fact, we have
shown above the simple systems are characterized with one-fold deformation
$(n=0)$ to be determined by the linear homogeneous function with the
self-similarity degree $q_0=1$ which corresponds to the real row of poles
(\ref{21eda}) with the logarithmically oscillating amplitude (\ref{abda}). The
passage to the complex systems caused by long-range interaction and long-time
memory arrives at creation of the first row of imaginary poles (\ref{21ed}).
These poles are related to the two-fold deformation $(n=1)$ characterizing the
Tsallis non-extensive thermostatistics with the self-similarity degree
$q_1(\lambda)$ whose value decays logarithmically slowly with dilatation growth
from the gold mean $q_1(1)\simeq 1.618$ to the minimum value $q_1(\infty)=1$.
As the amplitude (\ref{aada}) decays with the argument growth, the homogeneous
function (\ref{21a}) related to the Tsallis thermostatistics is written in the
form
\begin{equation}
h(x)=p_0 x+p_1\left(\frac{\ln x}{\ln\lambda_1}\right)x^Q,\quad Q\equiv
{q_1(\lambda_1)-\zeta_1}.
 \label{bbda}
\end{equation}
Here, we take into account that the lower degree $q_0=1$ and in the limit
$N\to\infty$ the row (\ref{21eda}) is reduced to the single pole $z_0^0=0$ to
transform the logarithmically periodical amplitude (\ref{abda}) into a constant
$p_0$ related to the limit dilatation $\lambda_0=N\to\infty$.

The equalities (\ref{bbda}) represent main finding of our investigation. It
means the homogeneous function of self-affine systems, being invariant under
both infinite $\lambda_0\to\infty$ and finite $\lambda_1\ne1$ dilatations, is
reduced to usual linear term accompanied with non-linear logarithmically
oscillating adding. The self-similarity degree $Q=Q(\lambda_1)$ of this
function is limited above by the gold mean to decay monotonically with the
dilatation $\lambda_1$ increase.

Finally, note the study of extremely complex systems whose behavior is governed
with more than one rows of imaginary poles (\ref{21ed}) is beyond the scope of
our consideration.

\end{document}